\documentclass[12pt, letterpaper, oneside]{article}

\usepackage{graphicx}
\usepackage[body={6.6in, 9in}, right=1in, top=1in]{geometry}
\usepackage{amssymb}
\usepackage{amsmath} 
\usepackage{color}
\usepackage[linktoc=page,colorlinks=false,linkcolor=cyan,citecolor=cyan,urlcolor=blue]{hyperref}
\usepackage{comment}
\usepackage{authblk}
\usepackage{subfig}

\def\be{\begin{equation}}
\def\ee{\end{equation}}
\def\ba{\begin{eqnarray}}
\def\ea{\end{eqnarray}}	
\def\l{\left}
\def\r{\right}
\def\fr{\frac}

\def\d{\partial}

\newcommand{\sfrac}[2]{{\textstyle\frac{#1}{#2}}}

\begin{document}



\begin{center}
\Large{\textbf{Scalar-tensor mixing from icosahedral inflation}} \\[1cm]
\large{Alberto Nicolis and Guanhao Sun}
\\[0.4cm]

\vspace{.2cm}
\small{\textit{Center for Theoretical Physics and Department of Physics, \\
  Columbia University, New York, NY 10027, USA}}

\end{center}

\vspace{.2cm}


\begin{abstract}
We study the mixed scalar-tensor two-point function in icosahedral inflation. Within the regime of validity of the effective field theory, this has to be perturbatively small; in particular, much smaller than the scalar spectrum. However, it can be much bigger that the tensor spectrum itself. We discuss observational implications for the CMB temperature-polarization spectrum.
\end{abstract}

\newcommand\numberthis{\addtocounter{equation}{1}\tag{\theequation}}
\newcommand\bi{\bibitem}
\newcommand\approxgt{\mbox{$^{>}\hspace{-0.24cm}_{\sim}$}}
\newcommand\approxlt{\mbox{$^{<}\hspace{-0.24cm}_{\sim}$}}
\newcommand\lsim{\mathrel{\rlap{\lower4pt\hbox{\hskip1pt$\sim$}}
        \raise1pt\hbox{$<$}}}
\newcommand\gsim{\mathrel{\rlap{\lower4pt\hbox{\hskip1pt$\sim$}}
        \raise1pt\hbox{$>$}}}
\newcommand\gm{{\rm geom}}
\newcommand\gw{{\rm grow}}

\pagestyle{plain}


\numberwithin{equation}{section}


\section{Introduction}
Cosmological observations indicate that the very early universe was approximately homogeneous and isotropic. However, the only primordial observables we have had access to so far are the background cosmology and the two-point function of scalar perturbations. Focusing on the observed isotropy, it is interesting to ponder whether the universe could secretely be anisotropic---in terms of other observables, such as higher-point correlation functions---but featuring an accidental isotropy for the observables we have detected so far. By `accidental' we mean something precise: that isotropy of those observables is a so-called accidental symmetry, akin to baryon number conservation in the standard model of electroweak interactions. That is, an approximate symmetry that is enforced by the fundamental symmetries of the theory to some low order in a perturbative expansion.
Icosahedral inflation is a concrete implementation of this idea \cite{icosahedralinflation}.

Icosahedral inflation can be thought of as inflation driven by a peculiar solid with icosahedral symmetry, which is a discrete subgroup of 3D rotations ($SO(3)$). In principle, the background cosmological evolution and all correlation functions for perturbations must be invariant under such discrete rotations, but not necessarily under generic continuous rotations. However, icosahedral rotations are so `dense' (in a colloquial sense) in $SO(3)$, that the background evolution and the scalar two-point function at long distances happen to be accidentally  isotropic \cite{icosahedralinflation}. Beyond these two observables, full isotropy is lost, and one can check explicitely that already the scalar three-point function and the tensor two-point function are generically anisotropic. In particular, the scalar three-point function can be maximally anisotropic \cite{icosahedralinflation}, i.e., can have vanishing overlap with all isotropic templates used in data analyses, and the tensor spectrum can have nonzero mixed correlators between the two helicities
\cite{tensortensor}.

Here, we show that the mixed scalar-tensor two-point function is also expected to be nonzero in icosahedral inflation. Such mixed correlator vanishes to lowest order in the derivative expansion. However, it is generically there once higher derivative corrections are taken into account. This makes it suppressed within the regime of validity of the derivative expansion, which is the relevant perturbative expansion for an effective field theory like ours. As a result, it is much smaller that the scalar spectrum. Still, since the tensor spectrum is also suppressed compared to the scalar one, there is a consistent choice of parameters that makes the scalar-tensor mixing more important than the tensor spectrum itself. Schematically,
\be
\frac{\langle \zeta \gamma \rangle}{\langle \zeta \zeta \rangle} \sim \Delta c^2_{\zeta \gamma}  \; ,
\qquad
\frac{\langle \gamma \gamma \rangle}{\langle \zeta \zeta \rangle} \sim \epsilon c_L^5 \; ,  
\ee
where $\Delta c^2_{\zeta \gamma}$ is a small dimensionless mixing parameter, $\epsilon = -\dot H/H^2$ is the usual slow-roll parameter, and $c_L$ is the propagation speed of scalar perturbations---which at short distances just reduce to longitudinal phonons, hence the `$L$'. One sees immediately that for $\epsilon c_L^5 \ll \Delta c^2_{\zeta \gamma}$, the mixed scalar-tensor correlator is bigger than the tensor spectrum.

\section{Icosahedral inflation}
%
%
%
%

Icosahedral inflation \cite{icosahedralinflation} is a variant of solid inflation \cite{solidinflation}. Apart from gravity, the degrees of freedom are
a triplet of scalar fields $\phi^I(x), I=1,2,3$, obeying shift symmetries and internal icosahedral rotation symmetries, 
\begin{equation} \label{internal}
\phi^I \rightarrow \phi^I+a^I, \;\;\; \phi^I = D^I {}_J\phi^J 
\end{equation}
where the $a^I$'s are constant shifts and $D$ is any element of the icosahedral group. To lowest order in derivatives, the basic building block for the action is  the matrix 
\begin{equation}
B^{IJ} = \partial_{\mu}\phi^I\partial^{\mu}\phi^J \; ,
\end{equation}
and, upon including gravity, the action reads
\begin{equation}
S_0 = \int d^4x \sqrt{-g} \Big[ \sfrac{1}{2}M_P^2 R + F \big( B^{IJ} \big) \Big] \; , 
\end{equation}
where $F$ is a generic function invariant under icosahedral rotations acting on the $I,J$ indices.

It can be checked \cite{icosahedralinflation, solidinflation} that such an action admits FRW solutions for the metric, with the scalar fields taking background values
\be \label{background}
\langle \phi^I \rangle = x^I \; ,
\ee
where the $x^I$'s are the usual FRW comoving coordinates. Moreover, such a solution describes an inflationary universe  with near exponential expansion if one demands that the action above further enjoy an approximate internal dilation symmetry, $\phi^I (x) \to  \lambda \, \phi^I(x)$ \cite{solidinflation}. (The slow-roll parameter $\epsilon$ can be thought of as a small breaking parameter for such an approximate symmetry.)

Considering all the spacetime and internal symmetries at our disposal, the FRW metric is invariant under spatial translations and rotations, and, trivially, under all transformations in \eqref{internal}. On the other hand, the scalar background configurations \eqref{background} are invariant under {\em (i)} the combined action of spatial translations and the internal shifts of \eqref{internal}, and {\em (ii)} the combined action of spatial icosahedral  rotations and  the internal icosahedral rotations of \eqref{internal}. So, overall, all background fields are invariant under {\em (i)} and {\em (ii)}, which, following standard spontaneously symmetry breaking (SSB) nomenclature, make up the unbroken subgroup, which has the algebra of 3D translations and icosahedral rotations.

When we introduce gravitational and matter perturbations,
\be
g_{\mu\nu} = g^{\rm FRW}_{\mu\nu} +h_{\mu\nu}\; , \qquad \phi^I = \langle \phi^I \rangle + \pi^I \; ,
\ee
their action will be manifestly invariant only under the unbroken subgroup. In particular, we can stop differentiating between spatial and internal indices, since they transform in the same way under the unbroken subgroup.
As usual, the broken symmetries are not lost---they are non-linearly realized on the perturbations---but we will not need them for the computations in this paper.

We refer the reader to the original papers \cite{icosahedralinflation, tensortensor, solidinflation} for more details about the general framework and the explicit construction of the model. 

\section{The mixed scalar-tensor two-point function} \label{perturbative}
In solid inflation, cosmological perturbations can be classified in terms of tensors ($\gamma_{ij}$), vectors/transverse phonons ($\vec \pi_T$), and scalars/longitudinal phonons ($\pi_L$) \cite{solidinflation}.
At the two-derivative level, after solving the constraints one finds the quadratic action
\be
S_{(2)} = S_{\gamma}+S_{L}+S_{T} \; ,
\ee
with \cite{solidinflation}
\begin{align}
S_\gamma &= \sfrac14 {M_{\rm Pl}^2} \int dt\,d^3x \,a^3\Big[\sfrac12 \dot{\gamma}_{ij}^2 -\sfrac{1}{2 a^2} \big(\d_m \gamma_{ij}\big)^2 +2\dot{H}c_T^2 \, \gamma_{ij}^2 \Big] \label{tensors}\\
S_{T} &=M_{\rm Pl}^2 \int dt \int_{\vec k} \,a^3 \bigg[\frac{ k^2/4}{1-k^2/4a^2\dot{H}} \, \big| \dot{\pi}_T^i \big|^2 +\dot{H}c_T^2 \, k^2 \big|\pi_T^i  \big|^2 \bigg] \label{vectors}\\
S_{L} &= M_{\rm Pl}^2 \int dt \int_{\vec k}  \, a^3 \bigg[ \frac{ k^2/3}{1-k^2/3a^2\dot{H}}\big|\dot{\pi}_L -({\dot{H}}/{H})\pi_L\big|^2+\dot{H}c_L^2 \, k^2 \big| \pi_L  \big|^2 \bigg] \; . \label{scalars}
\end{align}

For icosahedral inflation, since the  background does not have full $SO(3)$ symmetry, one expects quadratic mixings among these different polarizations---neither spin nor helicity are good quantum numbers. However, as pointed out already in \cite{icosahedralinflation, tensortensor}, such an effect is invisible to lowest order in the derivative expansion. On the other hand, if one takes into account higher derivative corrections, it is easy to write down mixing terms that are consistent with icosahedral symmetry. Ref.~\cite{tensortensor} considered the leading anisotropy effects for the tensor spectrum, which include a mixed correlator for helicities $+2$ and $-2$. Here we do the same for the scalar-tensor two-point function.

In a derivative expansion, the first icosahedral-invariant bilinear term we can write down that mixes scalars (and vectors) with tensors is
\begin{equation} \label{Smix}
S_{\rm mix} = -M_{\rm Pl}^{2}\int dt d^3x \, a \, {\Delta c_{\gamma\zeta}^{2}} \, T^{ijklmn}_{6} \partial_{i}\pi_{j}\partial_{k}\partial_{l}\gamma_{mn} \; .
\end{equation}
Here, $\Delta c^2_{\gamma\zeta}$ is a free dimensionless parameter---which we expect to depend slowly on time, but which we can take as constant to zeroth-order in the slow-roll expansion---and $T_6$ is the unique (up to normalization) spin-6 icosahedral invariant tensor \cite{tensortensor}.
As we show in Appendix \ref{interactionterm}, the single power of $a(t)$ is consistent with the near scale-invariance of the solid driving inflation, which is ultimately related to the slow-roll expansion \cite{solidinflation}. There, we also show that, to this order in derivatives, associated with \eqref{Smix} there are no extra scalar-tensor mixings involving $N$ or $N^i$. Finally, in the spirit of the derivative expansion and according to standard EFT logic, we need higher-derivative corrections to yield small effects at the scales of interest, that is, for typical frequencies of order $H$. This requires ${\Delta c_{\gamma\zeta}^{2}}$ to be generically `small'; how small will be made clear in sect.~\ref{nonperturbative}. 
The mixing term above  can come from non-minimal couplings between our solid and the Riemann tensor, e.g.~of the form $(R^{\mu\nu\rho\sigma}\partial_{\mu}\phi^I\partial_{\nu}\phi^J\partial_{\rho}\phi^K\partial_{\sigma}\phi^L)^3$ with suitable index contractions. We show this in the Appendix.

The fact that, within the regime of validity of the EFT, the term above can only yield small effects allows us to treat it in perturbation theory. 
Decomposing the phonon field into its longitudinal and transverse parts,
\be
\pi_{j} = \frac{\partial_{j}}{\sqrt{-\nabla^{2}}} \pi_{L} + \pi^{j}_{T} \; , \qquad \vec \nabla \cdot \vec \pi_T = 0 \; ,
\ee
and keeping only the longitudinal one, the mixing term above becomes
\begin{equation} \label{mixing}
S_{\rm mix} =  - M_{\rm Pl}^{2}\int \frac{d\tau d^{3}k}{(2\pi)^{3}} \, a^{2} \Delta c_{\zeta\gamma}^{2} \, T^{ijklmn}_{6} \, k_{i} \hat{k}_{j} k_{k}k_{l}  \, \pi_{L}(\vec{k}, \tau) \gamma_{mn}(-\vec{k},\tau) \; ,
\end{equation} 
where we switched to Fourier space and conformal time.
It is customary to parametrize scalar perturbations in terms of the variable $\zeta$, which for solid inflation is related to $\pi_L$ by  $\zeta = -k \, \pi_{L}/3$ \cite{solidinflation}. Following standard cosmological perturbation theory \cite{maldacena}, to first order in $\Delta c_{\zeta\gamma}^{2}$ the mixed two-point function we are after thus is
\begin{equation}
\langle\zeta(\vec k , \tau) \, \gamma^{s}(\vec q , \tau)\rangle = -i\int_{-\infty}^{\tau} d\tau' \langle\Omega(-\infty)|[\zeta(\vec k , \tau)\gamma^{s}(\vec q , \tau), H_{\rm int}(\tau')]|\Omega(-\infty)\rangle \;,
\end{equation}
where $s=\pm$ is either of the two tensor polarizations, 
\be \label{gamma ij}
\gamma_{ij}(\vec k,\tau) = \sum_{s=\pm} \gamma^s (\vec{k},\tau) \, \epsilon_{ij}^s(\vec k) \; , \quad \qquad \big( \epsilon^s_{ii} = k_i \epsilon^s_{ij} = 0 \, , \; \epsilon^s_{ij} \epsilon^{s'*}_{ij} = 2 \delta^{ss'}   \big) \; ,
\ee
and the interaction Hamiltonian is
\be
H_{\rm int}(\tau ') = -3M_{\rm Pl}^{2} \int \frac{d\tau ' d^{3}k' }{(2\pi)^{3}} a^{2} \, \Delta c_{\zeta\gamma}^{2} \,  T^{ijklmn}_{6} \, \hat{k}_{i}' \hat{k}_{j}' \hat{k}_{k}' \hat{k}_{l}' \, k'{} ^2\zeta(\vec{k}',\tau ')\gamma_{mn}(-\vec{k}',\tau ')
\ee

Writing our fields as usual as
\begin{align}
& \gamma^s (\vec{k},\tau) = \gamma_{cl}(k,\tau) \, a^s (\vec{k}) + \gamma^*_{cl}(k,\tau) \, a^{s\dagger}(-\vec{k}) \\
& \zeta(\vec{k},t) = \zeta_{cl}(\vec{k},t) \, b(\vec{k})  + \zeta_{cl}^*(\vec{k},t) \, b^{\dagger}(-\vec{k}) \; ,
\end{align}
and using the relevant mode functions to lowest order in slow roll \cite{solidinflation},
\begin{align}
& \gamma_{cl} (k,\tau) = \fr{1}{M_{\rm Pl} a}\,  \fr{e^{-ik\tau}}{\sqrt{k}} \,  \Big( 1 - \fr{i}{k\tau} \Big)  \\
& \zeta_{cl}(\vec{k},\tau) = -\fr{1}{M_{\rm Pl}a}\sqrt{\frac{c_{L}}{4\epsilon k}} {e^{-ikc_{L}\tau}} \Big(\frac{i}{c_{L}^{2}} + \frac{1}{c_{L}^{3}k\tau} + \frac{k}{3aHc_{L}} \Big) \; ,
\end{align}
after some straightforward (but tedious) algebra we get
\be \label{semifinal}
\langle\zeta \gamma^{s} \rangle' \equiv \frac{\langle\zeta ({\vec k , \tau}) \, \gamma^{s} (\vec q, \tau) \rangle }{(2\pi)^{3} \delta^{3}(\vec{k}+\vec{q})  }= \frac{\Delta c^{2}_{\zeta\gamma}}{\epsilon M_{\rm Pl}^{2}} \, T^{ijklmn}_{6} \, \hat{k}_{i} \hat{k}_{j} \hat{k}_{k} \hat{k}_{l}  \, \epsilon^{s}_{mn}(\vec{k}) \times I(\tau) \; ,
\ee
where
\begin{align}
I(\tau) \equiv \; & \frac32 \, \frac{c_{L}}{a^{2}} \Big[\Big(\frac{1}{c_{L}^{2}} - \frac{1}{k^{2}c_{L}^{3}\tau^{2}} + \frac{1}{3c_{L}} \Big) \frac{c_{L}^{2}+5c_{L}+3}{3kc_{L}^{2}(1+c_{L})^{2}} \nonumber \\
& + \Big(\frac{1}{kc_{L}^{2}\tau} + \frac{1}{kc_{L}^{3}\tau} + \frac{k}{3aHc_{L}} \Big)\Big(\frac{1}{k^{2}c_{L}^{3}\tau} - \frac{\tau}{3c_{L}(1+c_{L})}\Big)\Big] \; .
\end{align}

For late times, $k\tau \rightarrow 0^-$, our two-point function becomes time independent and scale invariant, and reduces to
\begin{equation}
\langle\zeta \gamma^{s} \rangle' =  \frac32 \, \frac{2c_{L}^{3}+4c_{L}^{2}+6c_{L}+3}{(1+c_{L})^{2}}  \cdot T^{ijklmn}_{6} \, \hat{k}_{i} \hat{k}_{j} \hat{k}_{k} \hat{k}_{l}  \, \epsilon^{s}_{mn}(\vec{k}) \cdot \frac{\Delta c^{2}_{\zeta\gamma}}{\epsilon c_L^5}\frac{H^{2} \, }{ M^{2}_{\rm Pl}k^{3}} \label{final}
\end{equation}
Dropping order-one numerical factors, 
\be
\langle \zeta \gamma  \rangle' \sim \frac{\Delta c^{2}_{\zeta\gamma}}{\epsilon c_L^5}\frac{H^{2} \, }{ M^{2}_{\rm Pl}k^{3}} \sim \frac{\Delta c^{2}_{\zeta\gamma}}{\epsilon c_L^5} \langle \gamma \gamma \rangle' \; ,
\ee
consistently with the estimate in \cite{tensortensor}, which was derived for $c_L \sim 1$. Recalling that for solid inflation models the 
 tensor-to-scalar ratio is roughly $r \sim \epsilon c_L^5$  \cite{solidinflation}, we see that for $\Delta c^{2}_{\zeta\gamma} \gg r$
the mixed correlator we computed is much bigger than the tensor spectrum itself. As we will see in sect.~\ref{nonperturbative}, such a possibility is still within the regime of validity of the effective theory and of a perturbative expansion in $\Delta c^{2}_{\zeta\gamma}$.

\section{Visualizing the two-point function}
The two-point function \eqref{final} depends on the orientation of $\vec k $ relative to the underlying icosahedral geometry, through the factor
\begin{equation} \label{M}
M^{\zeta s}(\vec k) \equiv T^{ijklmn}_{6} \, \hat{k}_{i} \hat{k}_{j} \hat{k}_{k} \hat{k}_{l}  \, \epsilon^{s}_{mn}(\vec{k})
\end{equation}
(we are using a notation consistent with that of \cite{tensortensor}, to facilitate comparison.)
As discussed at length  in \cite{tensortensor}, the phase of the polarization tensor $ \epsilon^{s}_{mn}(\vec k)$  is arbitrary, and, as a function of the direction of $\vec k$, necessarily involves singularities. This makes decomposing $M^{\zeta s}(\vec k)$ in spherical harmonics or plotting its angular dependence not particularly informative.

One possible way out is to consider the  squared absolute value of $M^{\zeta s}(\vec k)$, so that the ambiguous and singular phases cancel. Using the results of \cite{tensortensor},
\begin{align*}
|M^{\zeta+}|^{2} =|  M^{\zeta - }|^{2} & = \sfrac12 \sum_{s=\pm1}|M^{s\zeta}|^{2} \\
 &= \sfrac12 \, T^{ijklmn}_{6}T^{opqrst}_{6}\hat{k}_{i} \hat{k}_{j} \hat{k}_{k} \hat{k}_{l} \hat{k}_{o} \hat{k}_{p} \hat{k}_{q} \hat{k}_{r}\sum_{s=\pm1} \epsilon^{s}_{mn}(\vec{k})\epsilon^{s\;*}_{st}(\vec{k}) \\
&= \sfrac12 \, T^{ijklmn}_{6}T^{opqrst}_{6}\hat{k}_{i} \hat{k}_{j} \hat{k}_{k} \hat{k}_{l} \hat{k}_{o} \hat{k}_{p} \hat{k}_{q} \hat{k}_{r} (P_{ms}P_{nt}+P_{mt}P_{ns}-P_{mn}P_{st}) \numberthis \; ,
\end{align*}
where $P_{ij}$ is the transverse projector,
\begin{equation}
P_{ij}(\hat{k}) \equiv \delta_{ij}-\hat{k}_i \hat{k}_j \; .
\end{equation}

Following \cite{tensortensor}, we expect $|M^{\zeta s}|^{2} $ to contain spherical harmonics with $\ell = 0, 6, 10, 12$ only. Indeed, with the help of Mathematica we find
\be
|M^{\zeta s}|^{2} = \sfrac12 \, \sum_{\ell m} C_{\ell m}Y^{m}_{\ell}(\theta,\phi) \; ,
\ee
with the only nonzero $C_{\ell m}$ being
\begin{align*}
\mbox{$\ell = 0$:} \quad   & C_{0,0}=\sfrac{1024\sqrt{\pi}}{3003}(\gamma+1) \numberthis \\ 
\mbox{$\ell = 6$:} \quad  & C_{6,\pm6} = -\sqrt{\sfrac{5}{11}} \, C_{6,\pm2} = \sfrac{32}{323} \sqrt{\sfrac{11\pi}{273}}\,(\gamma+2) \\
& C_{6,\pm4} = -\sqrt{\sfrac{7}{2}} \, C_{6,0} = -\sfrac{352}{969} \sqrt{\sfrac{2\pi}{91}} \, (\gamma+1) \numberthis \\
\mbox{$\ell = 10$:} \quad  & C_{10,\pm10} = -\sqrt{\sfrac{255}{19}} \, C_{10,\pm6} = -\sqrt{\sfrac{255}{494}} \, C_{10,\pm2} = -\sfrac{20}{23} \sqrt{\sfrac{21\pi}{46189}} \, (3\gamma+1)\\
&C_{10,\pm8} = \sfrac{1}{2}\sqrt{\sfrac{17}{3}} \, C_{10,\pm4} = -\sqrt{\sfrac{187}{130}} \, C_{10,0} = -\sfrac{20}{23}\sqrt{\sfrac{70\pi}{7293}}  \,(\gamma+1) \numberthis \\
\mbox{$\ell = 12$:} \quad  & C_{12,\pm12} = 5\sqrt{\sfrac{69}{154}} \, C_{12,\pm8} = \sfrac{15}{17}\sqrt{\sfrac{437}{187}} \, C_{12,\pm4} = \sfrac{5}{58}\sqrt{\sfrac{5681}{119}} \, C_{12,0} = \sfrac{45}{2}\sqrt{\sfrac{\pi}{676039}} \, (\gamma+1) \\
&C_{12,\pm10} = -\sfrac{1}{5}\sqrt{\sfrac{209}{21}} \, C_{12,\pm6} = \sqrt{\sfrac{209}{34}} \, C_{12,\pm2} = \sfrac{33}{23}\sqrt{\sfrac{3\pi}{29393}} \, (3\gamma+1)  \; , \numberthis
\end{align*}
where $\gamma$ is the golden ratio.

\begin{figure}[h]
\centering
    \subfloat[$|M_{\zeta\gamma}|$ overlapping with our icosahedron]{{\includegraphics[scale=.61]{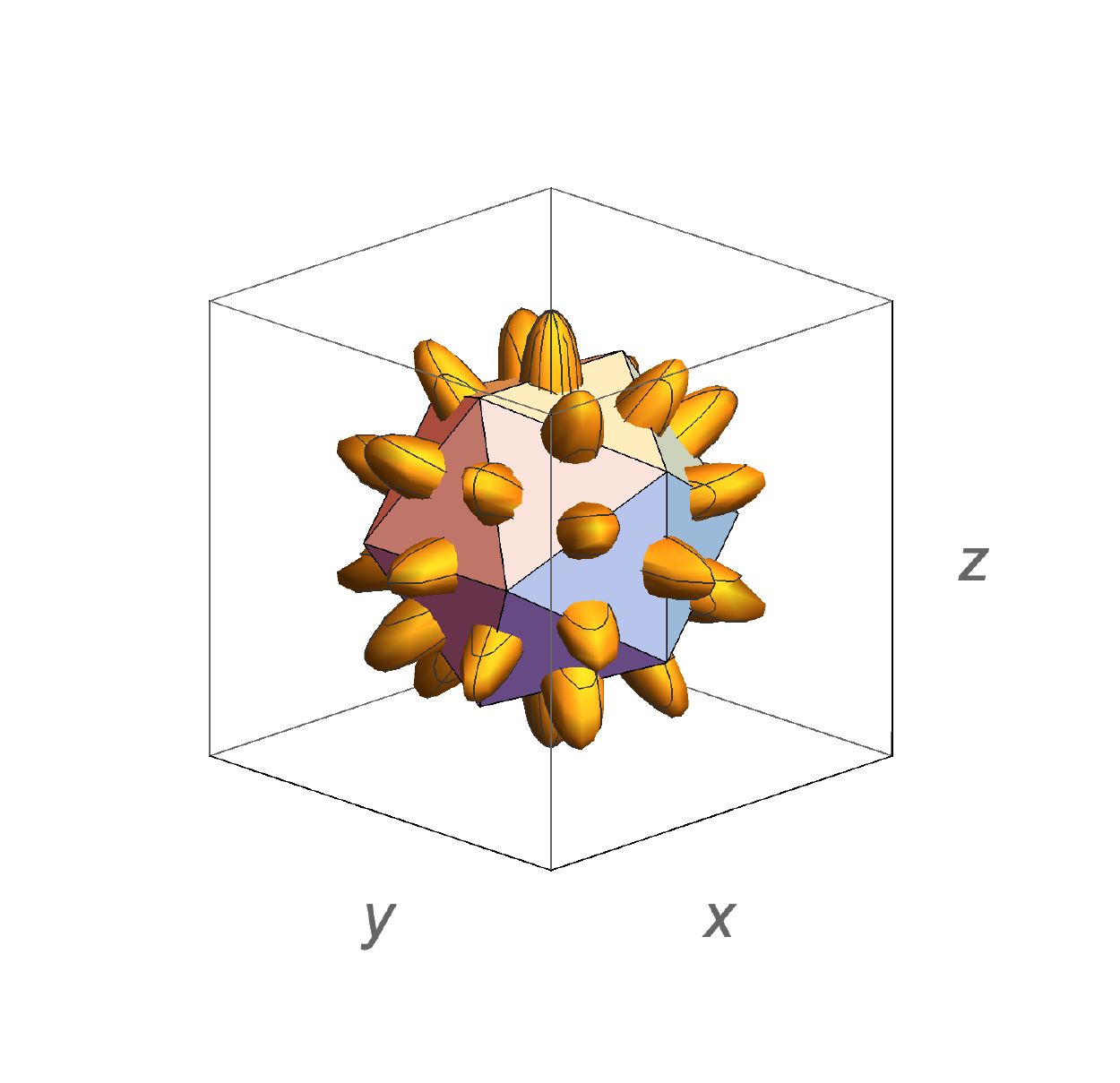} }}%
    \qquad
    \subfloat[$|M_{\zeta\gamma}|$ standing alone]{{\includegraphics[scale=.61]{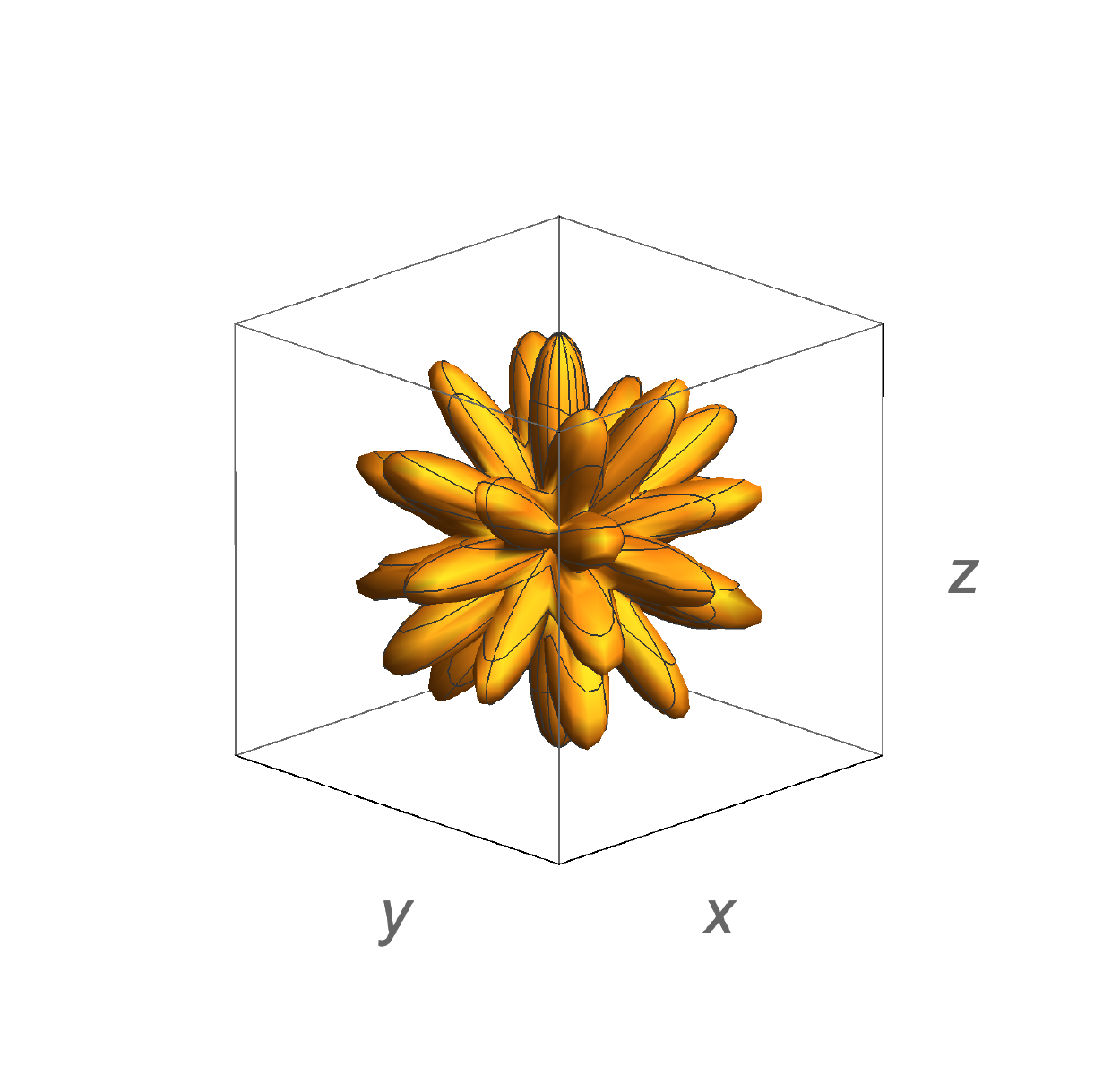} }}%
\caption{Angular plot of $|M_{\zeta\gamma}|$.\label{fig:aniso}}
\end{figure}

In Figure \ref{fig:aniso}  we plot the angular dependence of $|M_{\zeta\gamma}|$, alongside the underlying icosahedral structure. Clearly, the signal is concentrated around directions pointing towards the edges of the icosahedron. 

Another way to get rid of the ambiguous phases is to consider directly the two-point function $ \langle \zeta \gamma_{ij}\rangle$, because the full $\gamma_{ij}$ field---eq.~\eqref{gamma ij}---is unambiguous. Using the results above and the tracelessness of $T_6$, we get
\begin{align*}
\langle \zeta \gamma_{ij}\rangle & \propto \sum_{s=\pm1} M^{\zeta s }\epsilon^{s}_{ij}(-\vec{k})\\
& =T^{klmnop}_{6}\hat{k}_{k}\hat{k}_{l}\hat{k}_{m}\hat{k}_{n} \sum_{s=\pm1}\epsilon^{s}_{op}(\vec{k})\epsilon^{s\;*}_{ij}(\vec{k})\\
& = (P_{oi}P_{pj}+P_{oj}P_{pi}-P_{op}P_{ij}) \, T^{klmnop}_{6}\hat{k}_{k}\hat{k}_{l}\hat{k}_{m}\hat{k}_{n}\\
& = \big[ 2T_{6}^{klmnij}-2(T_{6}^{klmnip}\hat{k}_{p}\hat{k}_{j}+T_{6}^{klmnjp}\hat{k}_{i}\hat{k}_{p} ) +T_{6}^{klmnop}\hat{k}_{o}\hat{k}_{p}(\delta_{ij}+\hat{k}_{i}\hat{k}_{j}) \big] \hat{k}_{k}\hat{k}_{l}\hat{k}_{m}\hat{k}_{n} \numberthis \; .
\end{align*}
This however is a transverse traceless two-index tensor (because $\gamma_{ij}$ is), and so it is difficult to visualize its angular dependence: we cannot trace it or contract it with $\hat k$'s to construct a scalar angular function.

\section{Non-perturbative check}\label{nonperturbative}
As a check of our results of sect.~\ref{perturbative}, we now try to calculate the same two-point function in a non-perturbative way. We will be able do so only in a specific kinematical regime, which however will still allow us to perform a nontrivial check. 

Eventually we will still expand our result to linear order in $\Delta c^2_{\zeta \gamma}$, so, we can focus from the start on a two-field system made up of the scalar perturbations and either polarization of the tensor ones, because at linear order there cannot be interference among different sources of mixing. In particular, we can safely neglect the tensor-tensor mixing of ref.~\cite{tensortensor}.
Moreover, it turns out that, because of the time-dependence of $a(\tau)$, even this simple two-field system cannot be diagonalized for generic momenta (or generic times). We show this in Appendix \ref{diagonalappendix}. So, here we focus on modes well inside the sound horizon, $c_L k/aH \gg 1$, for which the time-dependence of $a(\tau)$ can be neglected.

With these qualifications in mind, to lowest order in slow-roll the quadratic action we need is (see eqs.~\eqref{tensors}, \eqref{scalars}, \eqref{mixing})
\begin{align}
S_{\gamma}+S_{L} +S_{\rm mix} & \to \sfrac{1}{2}M^{2}_{\rm Pl} a^2 \int \frac{d\tau d^{3}k}{(2\pi)^{3}} \Big[\sfrac{1}{2} \big(|\gamma_s'|^{2}  -  k^{2}|\gamma_s|^{2} \big) +2\epsilon a^{2}H^{2}(|\pi_{L}'|^{2}-c_{L}^{2}k^{2}|\pi_{L}|^{2})  \nonumber \\
& - \Delta c_{\zeta\gamma}^{2} k^3 \big( M^{\zeta s} \, \pi_L^* \gamma^s + {\rm c.c.} \big)\Big ]  \qquad \qquad \qquad (c_L k/aH \gg 1) \; ,
\end{align}
where $s$ is either $+$ or $-$, $M^{\zeta s}$ is defined in \eqref{M}, and all the fields and coefficients are evaluated at $(\vec k, \tau)$.
Neglecting the time-dependence of all the coefficients---including $a$---we can go to frequency space and rewrite this conveniently in a compact form as
\be
 \int \frac{d \omega d^3 k}{(2\pi)^4} \, \psi^{\dagger}  \cdot K \cdot \psi \; ,
\ee
where
\be
\psi \equiv 
\begin{pmatrix}
\gamma_{s}\\
\pi_L
\end{pmatrix} , \qquad
K \equiv
\sfrac{1}{2}M^{2}_{\rm Pl} a^2
\begin{pmatrix}
\sfrac{1}{2}(\omega^2-k^2)				& -\sfrac12 \Delta c_{\zeta\gamma}^{2} k^3 {M^{\zeta s}}^* \\
-\sfrac12 \Delta c_{\zeta\gamma}^{2} k^3 M^{\zeta s}		&  2\epsilon a^{2}H^{2} (\omega^2 - c_L^2 k^2)
\end{pmatrix},
\ee
$M^{\zeta s}$ is defined in \eqref{M}, and all the fields are now evaluated at $(\vec k, \omega)$.

To compute the equal-time two-point function we are interested in well inside the sound horizon, we can now simply invert the matrix $K$, insert the $i \epsilon$'s appropriate for the Feynman prescription for the poles, and take the integral over $\omega$ through standard residue methods. The reason this procedure is correct in our limit is that in general it gives the ground state's $T$-ordered correlation functions for a quantum system with a time-independent Hamiltonian; in our case, $T$-ordering does not matter, because our fields commute at equal time; moreover, in our inside-the-sound-horizon limit the time-dependence of the perturbations' Hamiltonian is negligible, and the Bunch-Davies ground state is equivalent to the flat-space one.

Then, the Fourier-space Feynman propagator of $\psi$  is schematically 
\be
\langle  \psi \psi^\dagger \rangle_{\omega, \vec k} = i     (K + i \epsilon) ^{-1} \, (2\pi)^4 \delta^{4} \; ,
\ee
and so the equal-time two-point function we are interested in is
\begin{align}
\langle  \pi_L \gamma^s \rangle'_{\vec k, \tau} & = \int \frac{d \omega}{(2\pi)} \, i (K + i \epsilon)_{12}^{-1} \nonumber \\
& =   \frac{\Delta c^{2}_{\zeta\gamma}}{ a^2  M_{\rm Pl}^{2} } k^3 M^{\zeta s}	 \int \frac{d \omega}{(2\pi)} \frac{i}{\epsilon a^2 H^2 (\omega^2-k^2 + i \epsilon)(\omega^2-c_L^2 k^2 + i\epsilon) - \sfrac14 \Delta c^{2}_{\zeta\gamma} |M^{\zeta s}|^2 k^6 } \nonumber \\
& \simeq - \frac{\Delta c^{2}_{\zeta\gamma}}{2 \epsilon H^2 a^4  M_{\rm Pl}^{2} } M^{\zeta s} \frac{1}{c_L(1+c_L)} \; ,
\end{align}
where in the last step we restricted to the first order in $\Delta c^{2}_{\zeta\gamma}$. Recalling that $\zeta$ is related to $\pi_L$  by $\zeta = - k\pi_L/3$, we see that this result matches precisely our previous one, eq.~\eqref{semifinal}, in the high $k$/early times limit, $c_L k |\tau| \gg 1$.

This computation also makes it clear how small $\Delta c^2_{\zeta\gamma}$ should be for a perturbative analysis to be applicable: the $\omega$ integral above is dominated by poles with $\omega \simeq \pm k$ and $\omega \simeq \pm c_L k$. The scalar-tensor mixing  shifts these, respectively, by
\be
\frac{\Delta \omega}{\omega} \simeq \frac{(\Delta c^{2}_{\zeta\gamma})^2 |M^{\zeta s}|^2}{8  (1-c_L^2)}  \frac{k^2}{\epsilon a^2 H^2 } \; , \qquad 
\frac{\Delta \omega}{\omega} \simeq - \frac{(\Delta c^{2}_{\zeta\gamma})^2 |M^{\zeta s}|^2}{8  c_L^2 (1-c_L^2)}  \frac{k^2}{\epsilon a^2 H^2 } \; .
\ee
For these relative shifts to be small up to physical momenta $k/a$ much bigger than $H$, we need 
\be
\Delta c^{2}_{\zeta\gamma} \ll c_L \sqrt{\epsilon} \; ,
\ee
where we used that  $M^{\zeta s}$ and $(1-c_L^2)$ are both of order one (in solid inflation models, $c_L^2$ has to be smaller than $1/3$ \cite{solidinflation}).

\section{Imprints on CMB Anisotropies}
We now turn our attention to the effects of a scalar-tensor mixing  on CMB anisotropies. In more standard cases, where the inflationary theory has both rotational and parity symmetry, a mixing between the so-called $E$ and $B$ modes is forbidden due to symmetry arguments. More precisely, following the convention in \cite{Weinberg:2008zzc}, 
\begin{align*}
\langle a^*_{T,lm} a_{T,l'm'} \rangle = &C_{TT,l} \, \delta_{l,l'} \, \delta_{m,m'}\\
\langle a^*_{T,lm} a_{E,l'm'} \rangle = &C_{TE,l} \, \delta_{l,l'} \, \delta_{m,m'}\\
\langle a^*_{E,lm} a_{E,l'm'} \rangle = &C_{EE,l} \, \delta_{l,l'} \, \delta_{m,m'}\\
\langle a^*_{B,lm} a_{B,l'm'} \rangle = &C_{BB,l} \, \delta_{l,l'} \, \delta_{m,m'}\\
\langle a^*_{T,lm} a_{B,l'm'} \rangle = &0\\
\langle a^*_{E,lm} a_{B,l'm'} \rangle = &0 \; .
\end{align*}
In particular, the $T$-$B$ and $E$-$B$ correlators vanish because under a parity transformation one has
\be
a_{T,lm} \rightarrow (-1)^l \, a_{T,lm} \;, \qquad a_{E,lm} \rightarrow (-1)^l \, a_{E,lm} \;, \qquad a_{B,lm} \rightarrow -(-1)^l \,  a_{B,lm} \; .
\ee
Therefore, when $l = l'$, such correlators are forbidden because of parity, while for $l \neq l'$, they are forbidden because of rotations. 

However, in our case there is no full rotational symmetry, hence modes of different $l$'s can mix. As a result, one can generically expect nonzero $\langle a^*_{T,lm} a_{B,l'm'} \rangle$ and $\langle a^*_{E,lm} a_{B,l'm'} \rangle$ correlators when $l = l' \pm n$ for odd $n$ (even $n$'s are still forbidden by parity, which is a symmetry of our theory). A similar argument has been presented in \cite{Bartolo:2014hwa} for pseudoscalar inflation. Adapting the notation of \cite{Bartolo:2014hwa} and \cite{Shiraishi:2010sm}, 
\begin{align}
a^{(s)}_{T/E, lm} =& 4\pi (-i)^l \int \frac{d^3 k}{(2\pi)^3} \mathcal{T}^{(s)}_{T/E,l}(k) \, \zeta_{\vec{k}} \, Y^{*}_{lm} (\hat{k})\\
a^{(t)}_{T/E, lm} =& 4\pi (-i)^l \int \frac{d^3 k}{(2\pi)^3} \mathcal{T}^{(t)}_{T/E,l}(k) \left[ \gamma_{\vec{k}}^{(+2)} {}_{-2} \, Y^*_{lm}(\hat{k}) + \gamma_{\vec{k}}^{(-2)} {}_{2} \, Y^*_{l\;m}(\hat{k})\right]\\
a^{(t)}_{B, lm} =& 4\pi (-i)^l \int \frac{d^3 k}{(2\pi)^3} \mathcal{T}^{(t)}_{B,l}(k) \left[ \gamma_{\vec{k}}^{(+2)} {}_{-2}\, Y^*_{lm}(\hat{k}) - \gamma_{\vec{k}}^{(-2)} {}_{2} \, Y^*_{l\;m}(\hat{k})\right]
\end{align}
Here we use $s$ and $t$ to label contributions from scalar and tensor modes, $\mathcal{T}^{(s/t)}_{T/E/B, l} (k)$ is the corresponding radiation transfer function (see, e.g., \cite{Shiraishi:2010sm} for their explicit forms), and $ {}_{\pm2}Y_{lm}$ is the spin-weighted spherical harmonics of spin $\pm 2$. The transfer functions depend only on the cosmology after inflation, and are thus independent of our inflationary model. Inflation enters the correlation functions above only through $\zeta$ and $\gamma$, evaluated at the end of inflation. (See \cite{Weinberg:2008zzc, Bartolo:2014hwa, Shiraishi:2010sm, Shiraishi:2010kd, Zaldarriaga:1996xe} for details). We thus have
\begin{align} 
&\langle a^{(s)*}_{T/E,lm} a^{(t)}_{B,l'm'} \rangle \nonumber \\
=&(4\pi)^2 i^{l-l'} (-1)^{l'} \times C \times A_{(l,m),(l',m')} \times \int \frac{dk}{k} \mathcal{T}^{(s)}_{T/E,l}(k) \mathcal{T}^{(t)}_{B,l'}(k)
\end{align}
where $C$ is a $k$-independent factor, given by our previous calculation as
$$
C = \frac{3}{2} \, \frac{2c^3_L+4c_L^2+6c_L+3}{(1+c_L)^2 } \, \frac{\Delta c^2_{\zeta\gamma} H^2}{\epsilon c_L^5   M^2_{Pl}} \; ,
$$
and $A_{(l,m),(l',m')}$ is a purely geometric factor, defined as
\be
A_{(l,m),(l',m')} \equiv \int d\Omega_{\hat{k}} \, Y_{lm}(\hat{k}) \left[M^{\zeta +} {}_{-2} \, Y^*_{l'm'}(\hat{k}) -  M^{\zeta -} {}_{2} \, Y^*_{l'\;m'}(\hat{k})\right] \;. 
\ee

For icosahedral inflation, the parity selection rules spelled out above allow non-vanishing $A_{(l,m),(l',m')}$ only for $ l = l' \pm n$, with $n$ odd.
In fact, further investigation with Mathematica  shows that there is no obvious selection rule based on the value of $l - l'$. 

Notice that the arbitrary and singular phase introduced in $M^{\zeta\pm}$ by the polarization tensors is still there---it does not cancel out in the combination entering $A_{(l,m),(l',m')}$. So, in order to evaluate these expressions, one should make an explicit choice of polarization tensors. For instance,  the choice of ref.~\cite{Shiraishi:2010kd} is,
\be
\epsilon_{mn}^{\pm 2}(\hat{k}) = \sqrt{2} \, \epsilon_m^{\pm 1}(\hat{k}) \, \epsilon_n^{\pm 1}(\hat{k}) \; , \qquad \epsilon_m^{\pm 1}(\hat{k}) = \frac{1}{\sqrt{2}} \big( \hat{\theta}(\hat{k}) \pm i \hat{\phi}(\hat{k}) \big) \; ,
\ee
where $\theta$ and $\phi$ are the polar and azimuthal angles of $\hat k$, and $\hat \theta$ and $\hat \phi$ are the corresponding unit vectors.
With this choice, as an example, for $l=3$, $l'=2$, $m'=2$, and arbitrary $m$, we find
\begin{align}
A_{(3,-2),(2,2)} =& \frac{2+\gamma}{3\sqrt{21}}\;,\nonumber\\
A_{(3,0),(2,2)} =& -\frac{\gamma}{6\sqrt{70}}\;,\nonumber\\
A_{(3,2),(2,2)} =& A_{(3,1),(2,2)} = A_{(3,-1),(2,2)} =0 \; ,
\end{align}
where $\gamma$ is, as before, the golden ratio (we computed the relevant integrals with Mathematica.)

Notice that icosahedral inflation also gives rise to nonzero tensor-tensor $T$-$B$ and $E$-$B$ correlators, $\langle a^{(t)*}_{T/E,lm} a^{(t)}_{B,l'm'} \rangle$, in addition to the scalar-tensor ones, $\langle a^{(s)*}_{T/E,lm} a^{(t)}_{B,l'm'} \rangle$. However,  $T$-$B$ and $E$-$B$ correlators are dominated by the latter contributions, since the anisotropies in the tensor-tensor spectrum are of order $\epsilon c_L^5 \ll 1$ compared to the tensor-scalar mixing \cite{tensortensor} \footnote{As shown in Appendix \ref{interactionterm}, the parameter $\Delta c^2_\gamma$ that corrects the tensor modes' propagation speed in an anisotropic fashion in ref.~\cite{tensortensor} is generically of the same order as our mixing parameter $\Delta c^2_{\zeta\gamma}$, since the two effects can arise from the same non-linear combinations of matter fields and curvature tensors.}.

Current CMB observations are able to put constraints on $T$-$B$ and $E$-$B$ correlations. In the CMB literature, such correlations are usually assumed to be coming from cosmic birefringence. For example, recent constraints on cosmic birefringence effect coming from ACT can be found in \cite{Namikawa:2020ffr} and \cite{Choi:2020ccd}, and similar constraints from Planck can be found in \cite{Aghanim:2016fhp} and \cite{Gruppuso:2020kfy}. However, it is not straightforward to translate constraints on cosmic birefringence into constraints on the parameters of our model. We leave performing this analysis for future work.

\section{Concluding remarks}

We have computed the scalar-tensor correlation function in icosahedral inflation \cite{icosahedralinflation}, and discussed its possible imprints on CMB anisotropies, in the form of non-vanishing $T$-$E$ and $T$-$B$ spectra. Such correlations are allowed because the inflationary model at hand breaks (spontaneously) rotational invariance. Within the regime of validity of the effective field theory, the mixed scalar-tensor correlator can be parametrically larger that the tensor spectrum itself. 

It is useful to compare our results and framework to other models of inflation featuring anisotropic effects, such as the model studied in \cite{Watanabe:2009ct, Gumrukcuoglu:2010yc, Watanabe:2010bu}. There, the intrinsic anisotropy of the background evolution enters all correlators of perturbations, including the scalar spectrum. In our case instead, the model is {\em designed} in such a way as to guarantee that the scalar spectrum is automatically isotropic, while leaving open the possibility of detectable anisotropies in other correlation functions, such as the scalar three-point function \cite{icosahedralinflation}, the tensor spectrum \cite{tensortensor}, and the scalar-tensor two-point function (the case considered here). The reason behind this choice is spelled out in the Introduction: the scalar spectrum is the only primordial correlation function we have detected, and it appears to be consistent with statistical isotropy.

%

\section*{Acknowledgements}
We thank Colin Hill and Lam Hui for useful discussions and comments.
Our work is partially supported by the US DOE (award number DE-SC011941) and by the Simons Foundation (award number 658906).

\section*{Appendix}
\appendix
\section{Origin of the mixing term and powers of $a$}
\label{interactionterm}

As emphasized in \cite{icosahedralinflation, tensortensor}, for icosahedral inflation anisotropies in two-point functions can only arise from higher derivative corrections to the solid's action. A possible candidate is a term schematically of the form $T_6 \cdot (R^{\mu\nu\rho\sigma}\partial_{\mu}\phi^{I}\partial_{\nu}\phi^{J}\partial_{\rho}\phi^{K}\partial_{\sigma}\phi^{L})^{3}$, with suitable $I$-type index contractions.
In fact, for such a term to be compatible with the approximate internal scale invariance associated with slow-roll \cite{solidinflation}, 
\be
\phi^I \to \lambda \phi^I \; ,
\ee
we need to multiply it by a factor scaling like $(B^{IJ})^{-6}$ (to lowest order in slow roll), again with suitable index contractions.

So, let's consider a higher-derivative action term schematically of the form
\be \label{DeltaS}
\Delta S  \sim \frac{1}{M^2} \int d\tau d^3 x \, a^4   \, (B^{IJ})^{-6} \, T_6 \cdot (R^{\mu\nu\rho\sigma}\partial_{\mu}\phi^{I}\partial_{\nu}\phi^{J}\partial_{\rho}\phi^{K}\partial_{\sigma}\phi^{L})^{3} \; ,
\ee
where we introduced an arbitrary dimensionful coupling constant.
Notice that we are using directly conformal time, since it makes the analysis that follows simpler: all  components of the unperturbed metric scale like $a^2$, and so we don't need to differentiate between time and space.

Consider now expanding such a term in spatially flat slicing gauge about our inflationary background. We are only interested in bilinear scalar-mixing terms. We expand our fields  in perturbations,
\be
\phi \to  x + \pi \; , \qquad g \to a^2(\tau) (\eta + h) \; ,
\ee
where $g$ is shorthand for the metric (with lower indices), and keep in mind that $h_{\mu\nu}$ contains tensors ($\gamma_{ij}$) as well as scalars ($h_{00}$ and $h_{0i} = \d_i \psi$). We thus need to keep $\pi$-$h$ and $h$-$h$ bilinear terms. However,  to this order in derivatives, the latter cannot contribute to our scalar-tensor mixing. The reason is that for the contraction with $T_6$ in \eqref{DeltaS} to yield something nonzero, we need six free spatial indices on the fields and their derivatives. If we neglect $\pi$, the factors of $\d \phi$ and $B^{IJ}$ yield terms with no derivatives on the fields, whereas the factors of $R^{\mu\nu\rho\sigma}$  yield terms with zero, one, or two-derivative per field. So, among the $h$-$h$ terms, the only ones that have a chance of giving a contribution to scalar-tensor mixings are of the form
\begin{align}
& \d_i \d_j \gamma_{kl}  \times \big( \d_m \d_n h_{00} \, ,  \; \d_m h_{0n} \, ,  \; \mbox{or } \d_m \d_0 h_{0n} \big) \\
& \big( \d_0 \d_i \gamma_{jk}  \, ,  \; \mbox{or } \d_i \gamma_{jk} \big)   \times  \d_l \d_m h_{0n}  \; ,
\end{align}
and all come from two factors of $R^{\mu\nu\rho\sigma}$ each expanded to linear order in $h_{\mu\nu}$. But then these terms cannot be there: the spacetime indices of the Riemann tensor are contracted with factors of $\d \phi$; if we neglect $\pi$, $\d_\mu \phi^I = \delta_\mu^I$, which projects all the indices of the Riemann tensor onto spatial directions; to linear order in the metric fluctuations, this can only yield spatial derivatives of $\gamma_{ij}$ and, in particular, no term involving $h_{00}$ or $h_{0i}$.   

So, in summary, to figure out our scalar-mixing terms in spatially flat-slicing gauge, we only need to look for $\pi$-$\gamma$ terms with three derivatives overall, and we can neglect everything else.
From the expansion of $\d \phi$ and $B^{IJ}$ we get first derivatives of $\pi$,
\be
\d \phi \to 1 + \d \pi \; , \qquad B^{IJ} \sim g^{-1} \d \phi \d\phi \to  \frac{1}{a^2} \big(1 + \d \pi  \big) \; ,
\ee
and from the Riemann tensor we get second derivatives of $\gamma$,
\begin{align}
R^{\mu\nu\rho\sigma} & \sim (g^{-1} )^3 R^{\mu} {}_{\nu\rho\sigma} \\
& \sim (g^{-1} )^3 \big( \d \Gamma + \Gamma \Gamma \big) \\
& \sim (g^{-1} )^5 \big( \d g \d g + g \, \d^2 g \big) \to  \frac{1}{a^6} \big(a^2 H^2 + \d^2 \gamma \big)\; ,
\end{align}
where we used that the Christoffel symbols are schematically $\Gamma \sim g^{-1 } \d g$ and that $\d_\tau a = a^2 H$.

Plugging all this into \eqref{DeltaS}, and keeping only the bilinear $\pi$-$\gamma$ terms, we get
\begin{align}
\Delta S  & \sim \frac{1}{M^2} \int d\tau d^3 x \, a^4 a^{12} (1+ \d \pi)^{-6} \, T_6 \, a^{-18}\big(a^2 H^2 + \d^2 \gamma \big)^3 (1+ \d \pi)^{12} \\
& \to \frac{H^4}{M^2} \int d\tau d^3 x \, a^2 \, T_6 \, \d \pi \, \d^2\gamma \; ,\label{dpiddh}
\end{align}
which, if written in proper time $t$, is precisely of the form \eqref{Smix}, with
\be
\Delta c^{2}_{\zeta\gamma} \sim \frac{H^4}{M^2 M_{\rm Pl}^2} \; ,
\ee
which is of the same order as the tensor-tensor mixing parameter $\Delta c^{2}_{\gamma}$ of ref.~\cite{tensortensor}.
Notice that, given the tracelessness and total symmetry of $T_6$, the index contraction displayed in \eqref{Smix} is the only non-vanishing one.

More in general, we can prove that the mixing term \eqref{Smix} is the only scalar-tensor bilinear mixing allowed by gauge invariance. To minimize the number of extra terms coming from covariant derivatives, here we use cosmic time $t$ instead of conformal time $\tau$, and $h$ now stands for metric perturbations about the FRW metric,
\be
g = g_{\rm FRW} + h \; .
\ee

First, let's set the fluctuations of $\phi^I$ to zero by fixing the spatial diffs (unitary gauge),
\be
\phi^I = x^I
\ee
In terms of gauge transformations $x^{\mu} \rightarrow x^{\mu} + \xi^{\mu}$, we have fixed the spatial components $\xi^i(x)$. Under the residual $\xi^0(x)$ transformations, $h$ transforms as
\be
h_{00} \rightarrow h_{00} + \d_0 \xi_0 \; , \qquad h_{0i} \rightarrow h_{0i} + \d_i \xi_0 \; , \qquad h_{ij} \rightarrow h_{ij} -2\delta_{ij}\dot{a} a \xi_0 \;.
\ee
The gauge-invariant building blocks for a possible mixing term thus are
\be
\quad \d_0 h_{0i} - \d_i h_{00}  \qquad \mbox{and} \qquad \d_k h_{ij} + \delta_{ij}\d_0\l( a^2 h_{0k} \r) - a^2\delta_{ij}\d_k h_{00}  \; .
\ee
We can now play the Stueckelberg trick and restore gauge invariance under the broken diffeomorphism $x^{i} \to x^{i} + \xi^{i}$  by promoting the parameters $\xi^i$ to Goldstone fields: 
\be
\xi^i(x) = \pi^i(x) \; .
\ee
Our building blocks, which are now invariant under all gauge transformations, now read
\begin{align}
\Pi_i &=\d_0 h_{0i} - \d_i h_{00} + \d_0\l(a^2\d_0\pi_i \r) \\
\Xi_{ijk} &= \d_k h_{ij} + \delta_{ij}\d_0\l( a^2 h_{0k} \r) - a^2\delta_{ij}\d_k h_{00} + a^2 \d_k \l( \d_i\pi_j + \d_j \pi_i \r) + \delta_{ij}\d_0 \l( a^4 \d_0 \pi_k \r)\; .
\end{align} 
Up to integration by parts, the combinations of these building blocks that may give rise to scalar-tensor mixing terms, with the least number of derivatives and correct number of free indices (six), are 
\be
\d_i \Pi_j \, \d_k \Xi_{mnl} \quad \mbox{and} \quad \Xi_{ijk} \, \Xi_{mnl}
\ee
The first term is suppressed in general because it has more derivatives than the second term. The  second term produces our $\d\pi\d^2\gamma$ mixing, the one appearing in \eqref{Smix}. Notice that the terms proportional to Kronecker deltas will not contribute to the final result since our icosahedral spin-6 tensor $T_6$ is traceless.

\section{Diagonalizability}
\label{diagonalappendix}

Consider the scalar-tensor sector of our theory, defined by the quadratic action terms \eqref{tensors}, \eqref{scalars}, and \eqref{mixing}. For simplicity, let's focus on a two-dimensional field space, made up of the scalars and a single polarization of the tensors (cf.~sect.~\ref{nonperturbative}). We want to see under what conditions such a system is diagonalizable. It is easier to work directly with the equations of motion rather than the action, since for the former there are no integration-by-parts ambiguities. Certainly, if the equations of motion cannot be made diagonal, neither can the action.

At fixed $\vec k$, upon changing the normalization of $\pi_L$ in a suitable time-dependent fashion, our equation of motion can be written compactly as
 \be
 \alpha \phi''+ \alpha' \phi'+ \alpha M \phi=0 \; ,
 \ee
 where $\phi$ is a doublet of fields, $\alpha$ is a scalar function of time, and $M$ is a $2\times2$ 
matrix, also time-dependent. 
 
 Let us assume that we can diagonalize such an eom through some time-dependent invertible matrix $R(\tau)$. By plugging $\phi = R(\tau)\varphi$ into the equation above, we get 
 \be
 \varphi''+\big(2R^{-1}R'+\sfrac{\alpha'}{\alpha}\big)\varphi'+\big(R^{-1}R''+\sfrac{\alpha'}{\alpha}R^{-1}R'+R^{-1}MR \big)\varphi=0 \; .
 \ee
For this equation to be diagonal, we need:
\begin{enumerate}
\item $R^{-1}R'$ to be diagonal, because of the $\varphi'$ term. However, this implies that $R^{-1}R''$ is also diagonal. This follows from
\be
R^{-1}R'' = \big(R^{-1}R' \big)' - (R^{-1})'R' =  \big(R^{-1}R' \big)'  + (R^{-1}R')^{2} \; ,
\ee
where the last equality can be got from taking the time derivative of $R^{-1}R=1$.

\item $R^{-1}MR$ to be diagonal, because all the other terms multiplying $\varphi$  already are.

 
 \end{enumerate}
But for $R^{-1}MR$ and $R^{-1}R'$ to be diagonal at the same time, we need them to commute. 
This in turn implies that $M$ commutes with $M'$. The reason is that if $R^{-1}MR$ is diagonal at all times, so is its time derivative, which is simply
 \be
 (R^{-1}MR)'= R^{-1}M'R+[R^{-1}MR, R^{-1}R'] = R^{-1}M'R \; .
 \ee
 So, $R^{-1}M'R$ is diagonal, which means that it commutes with $R^{-1}M R$. Or, equivalently, 
\be
[M', M]=0 \; .
\ee
This is a nontrivial condition on the time-dependence of $M$. It is easy to check that such a condition is not satisfied in our case. Therefore, our equation of motion is not diagonalizable. 
 
%
%
%
%

\end{document}